<u>Measuring Tempo, Rhythm, Timing,</u>

<u>and the Torques that Generate Power in the Golf Swing</u>

Robert D. Grober
Department of Applied Physics
Yale University
New Haven, CT  06520

January 4, 2010

This paper summarizes two previously published technical papers describing what can be learned from measuring the motion of the golf club using two accelerometers. The sensors are mounted at different points along the shaft of the golf club, both sensitive to acceleration along the axis of the shaft. Two distinct signals are resolved. The first signal scales as club speed, which can be used to understand details of tempo, rhythm, and timing. The second measures the acceleration of the hands, which reveals the torques that generate power in the golf swing. A comparative study of twenty-five golfers shows that club speed is generated in the downswing as a two step process. The first phase starts at the top of the swing and involves impulsive acceleration of the hands and club. This is followed by a second phase, the release, where the club is accelerated while the hands decelerate.

Introduction

This review summarizes two previously published technical papers [1, 2] describing what can be learned by measuring the motion of the golf club using two accelerometers. The sensors are mounted in the shaft of a golf club with the sensing direction oriented along the axis of the shaft. One accelerometer is located under the grip, preferably at a point between the two hands. The other is located further down the shaft. The output of the accelerometers is digitized and broadcast wirelessly to a computer, enabling data storage and signal analysis.

Differential Mode Signal

The data is resolved into two signals. The first, the differential mode signal, is a measure of the centripetal acceleration of the club. Because centripetal acceleration scales as the square of the velocity, it is a very reasonable proxy for the speed of the club. As a result the differential mode signal can be used to provide insight into the tempo, rhythm, and timing of the golf swing. Its simplicity enables real-time biofeedback and is the foundation for the instrumentation sold by Sonic Golf, Inc. as System-I [3].

Shown in Fig. 1 is the differential mode signal averaged over seven swings of a professional golfer. Because the differential mode signal is a direct measure of the speed of the club, there are several points in the swing which are easily identified. They include 1) the beginning of the swing (red circle), 2) point of maximum speed in the backswing (green circle), 3) transition from backswing to downswing (blue circle), and 4) impact (black circle). Note that impact results in a large shock to the system, which helps to identify the exact location of impact. When averaged over several swings, this shock appears as the anomaly seen just to the right of the black circle in Fig. 1.

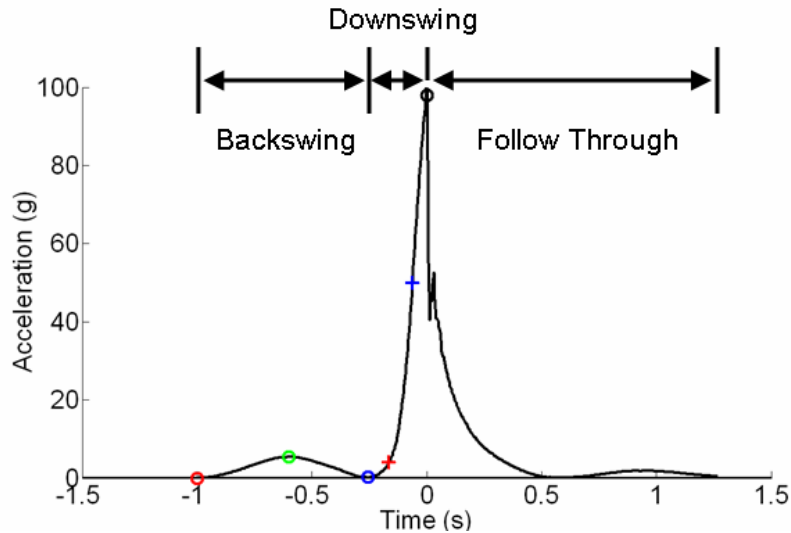

Fig 1. The differential mode signal averaged over seven swings of a professional golfer. The signal is representation of the speed of the club. Impact is centered at $t = 0$. The red circle marks the start of the swing; green circle is the peak speed during the backswing; blue circle is the transition from backswing to downswing; and the black circle indicates impact. The red and blue crosses are defined in Fig. 2. The anomaly just to the right of impact is due to the shock of impact.

As an example of the usefulness of this signal, one can determine the duration of the backswing and downswing for each individual swing, and then calculate the mean and standard deviation of the entire data set. For the data of Fig. 1, the average duration of the backswing is $731 \pm 21$ ms, and the average duration of the downswing is $258 \pm 8$ ms. The resulting ratio of backswing to downswing time is 2.8:1. This is very nearly the ratio of 3:1 first discussed by John Novosel in the book *Tour Tempo* [4], and subsequently by Grober, *et al*. [5], in which Novosel's observations were confirmed and a biomechanical explanation for the ratio was hypothesized. Note that these measurements and the corresponding analysis can be made instantaneously for every swing.

Common Mode Signal

The second signal derived from the data, the common mode signal, is a measure of the acceleration of the hands. Actually, it is a measure of the acceleration of the hands only as long as the wrists are cocked. This is the situation for just about just about all of the downswing, where the signal is most interesting. Acceleration only happens in response to applied force, and so this signal provides insight into the torques that generate speed during the downswing.

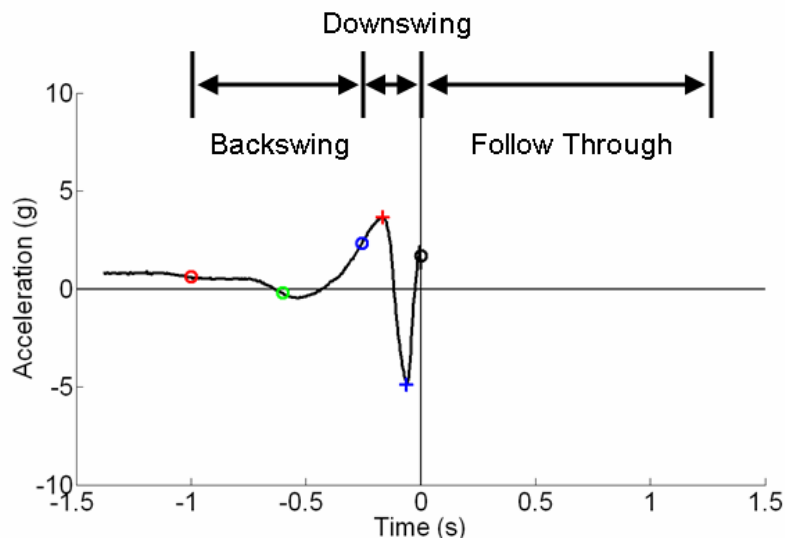

Fig 2. The common mode signal averaged over seven swings of a professional golfer. Impact is centered at $t = 0$. The interesting structure in this signal occurs during the downswing and is described in the text. The red cross marks the peak and the blue cross marks the minimum of this structure. The open circles were obtained from Fig. 1 and mark key points in the swing. We show the signal only until impact. The signal immediately after impact is dominated by the shock of impact and is not particularly relevant to this discussion.

Shown in Fig. 2 is the common mode signal calculated for the same data set shown in Fig. 1. As before, the data is the averaged over seven swings of a professional golfer. The four points measured in Fig. 1 (i.e. red, green, blue, and black circles) are indicated in Fig. 2 so that the key points of the swing are easily identified. The data is only graphed thru impact, as the region just after impact is dominated by the shock of impact and yields no particularly useful information. The most interesting feature in this common mode signal is the structure observed during the downswing, characterized by the local maximum (red cross), which occurs very near the beginning of the downswing, and the local minimum (blue cross), which occurs about half way to impact. The position of both these features relative to the differential mode signal is indicated in Fig 1.

As was stated above, during the downswing the common mode signal measures the acceleration of the hands. Thus, the peak near the red cross corresponds to a region of rapid acceleration of the hands (i.e. the hands are speeding up), and the negative valued minimum near the blue cross corresponds to a deceleration of the hands (i.e. the hands are slowing down).

To understand why the hands accelerate and decelerate it is necessary to consider the applied torque. A detailed, technical discussion is presented in Ref [2]. In the spirit of this paper, the following couple of paragraphs serve as a non-technical summary. The interested reader is encouraged to read the detailed discussion provided in Ref [2].

The torque that accelerates the hands at the beginning of the downswing is the torque generated by the body as the downswing begins. *The greater this applied torque, the greater the acceleration of the hands and club, and the larger is this positive going region of the common mode signal*.

Two torques work to decelerate the hands during the downswing. The first is the torque which the golfer applies to release the club (i.e. un-cock the wrists). This torque serves to accelerate the club while simultaneously decelerating the hands. The second torque is related to the centripetal acceleration of the club, which is what is measured in Fig. 1 (i.e. the differential mode signal). The faster the club speed, the larger the signal in Fig. 1, and the more the hands decelerate.

In summary, the negative going region of the common mode signal is a direct measure of the deceleration of the hands. *The greater the torque that releases the club and the faster the club moves, the more the hands decelerate and the larger the negative going dip in the common mode signal*.

Eventually the common mode signal goes through a local minimum. This generally occurs one-half to two-thirds of the way from the beginning of the downswing to impact, as can be seen by the position of the blue cross in Fig. 1. After this point, the common mode signal increases, eventually becoming positive very near to impact. The wrists fully un-cock during this later region of the swing and the common mode signal no longer measures the acceleration of the hands. What it does measure is a parameter with a positive value. Thus, the return of the common mode signal to a positive number very near to impact is a measure of the hands unhinging.

In summary, the common mode signal typical of the golf swing of professional golfers has a distinctive structure during the downswing. It exhibits a positive valued maximum followed by a negative valued minimum, and finally becomes slightly positive just as the club gets to impact. The positive maximum is associated with the impulsive acceleration of the hands and club at the beginning of the downswing. The negative minimum is associated with the release of the club, which serves to decelerate the hands. *The greater the common mode maximum, the larger is the torque that accelerates the*

*hands and club at the start of the downswing. The deeper the common mode minimum, the larger are the torques that release of the club.*

Detailed Comparison of Five Golfers

Figs. 3 thru 7 show the differential and common mode signals for five very different golfers. The first is an elderly, avid, amateur golfer with a handicap of order twenty-five, the second is an avid, amateur golfer with a handicap of order ten, the third is a competitive collegiate golfer, the fourth is a PGA tour professional, and the fifth is a professional long drive competitor. In each case it is worth comparing the peak of the differential mode signal, which provides an indication of how fast the club is moving at impact, with the size of the max-min structure in the common mode signal, which provides some indication for how club speed is developed. Commentary for each data set is provided in the figure captions. These figures are meant to provide some insight as to how the data evolves from the case of an avid golfer who does not hit the ball very far to the case of a golfer who hits the ball well over 300 yards.

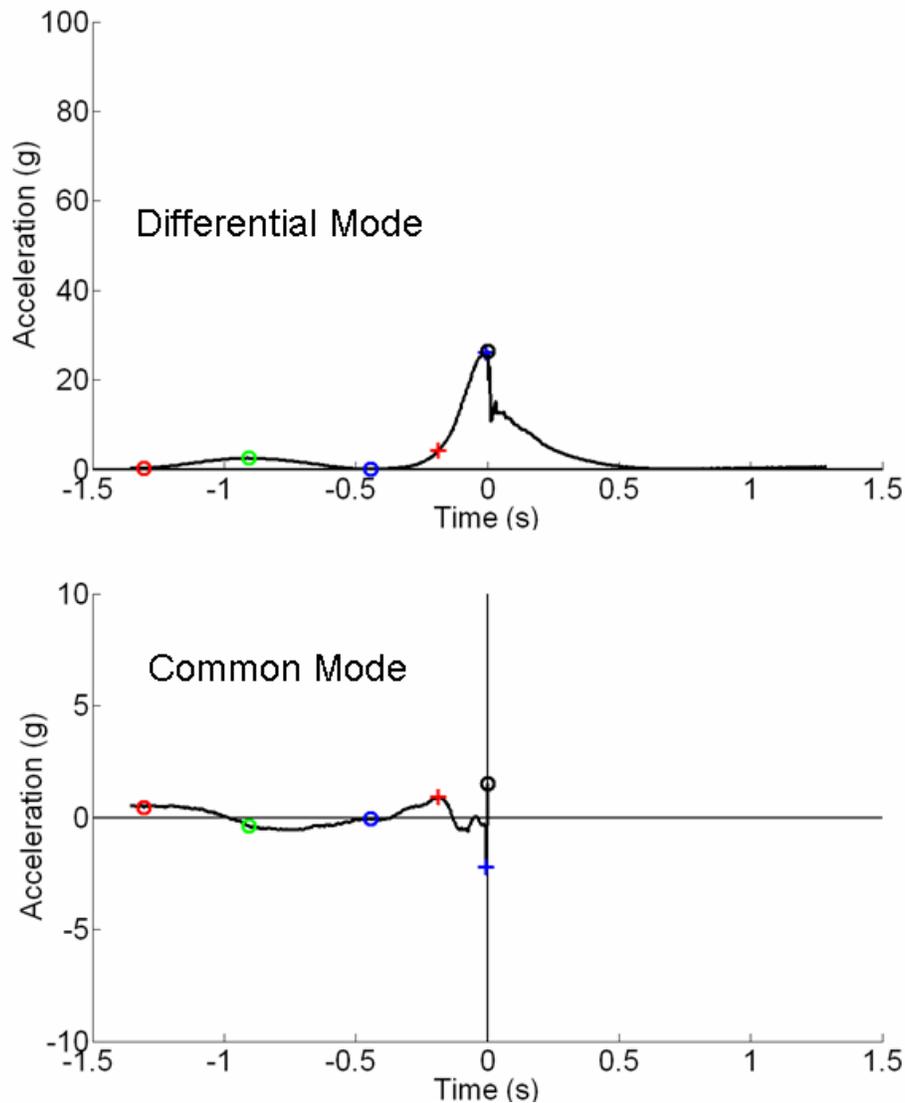

Fig. 3: Differential and common mode data averaged over seven swings of an elderly, avid, amateur golfer with a handicap of order twenty-five. The duration of the backswing is 830 ± 18 ms and the duration of the downswing is 433 ± 18 ms, which is remarkably consistent and characteristic of this beautifully rhythmic golf swing. However, this golfer is not particularly strong, as is clear from the peak of the differential mode and the very small structure in the common mode. It is interesting that while the duration of the backswing is only slightly slower than that of the professional golfer (Fig. 6), the duration of the downswing is much slower, suggesting significantly less physical strength.

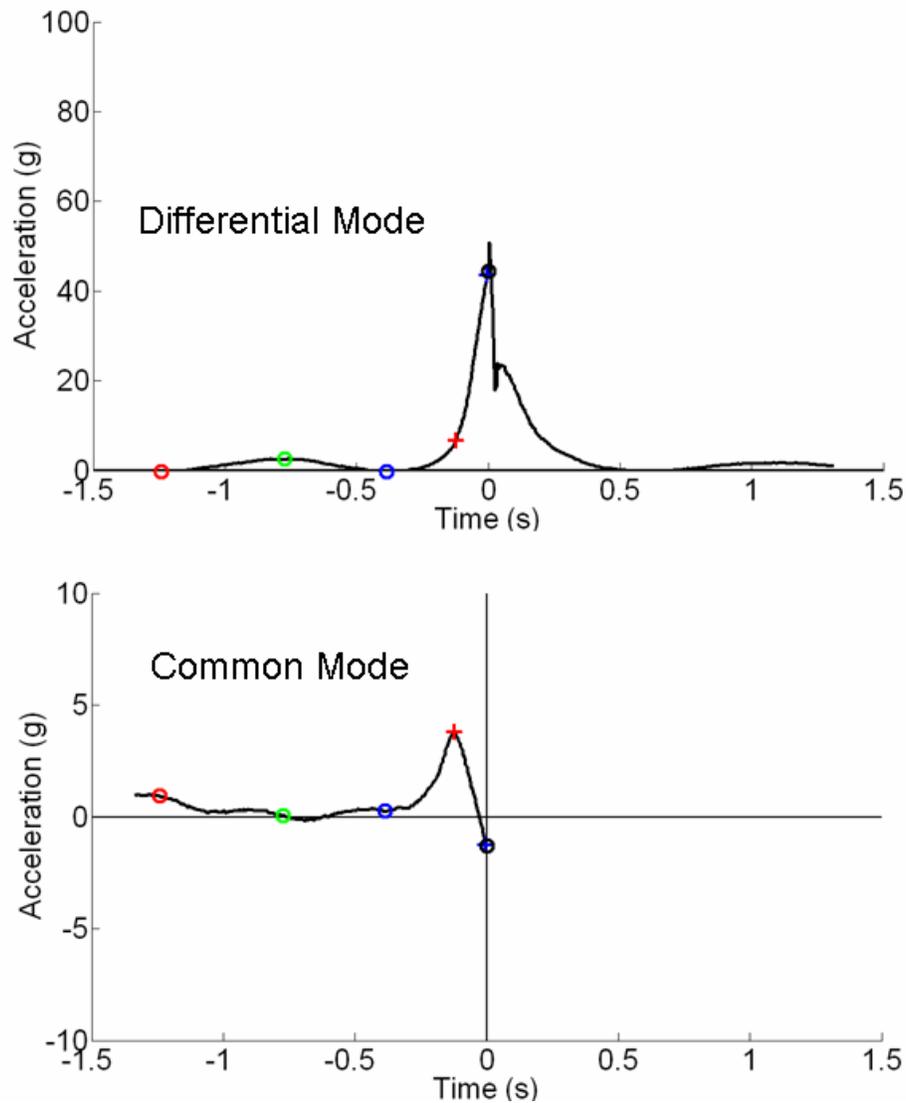

Fig. 4: Differential and common mode data averaged over six swings of an amateur golfer with a handicap of order ten. The duration of the backswing is 817 ± 16 ms and the duration of the downswing is 385 ± 20 ms. It is particularly interesting to note that while this golfer has a very respectable acceleration of the hands at the beginning of the downswing, the club never fully releases. This is generally done so as to obtain better control over the club head, but at the cost of significant club speed.

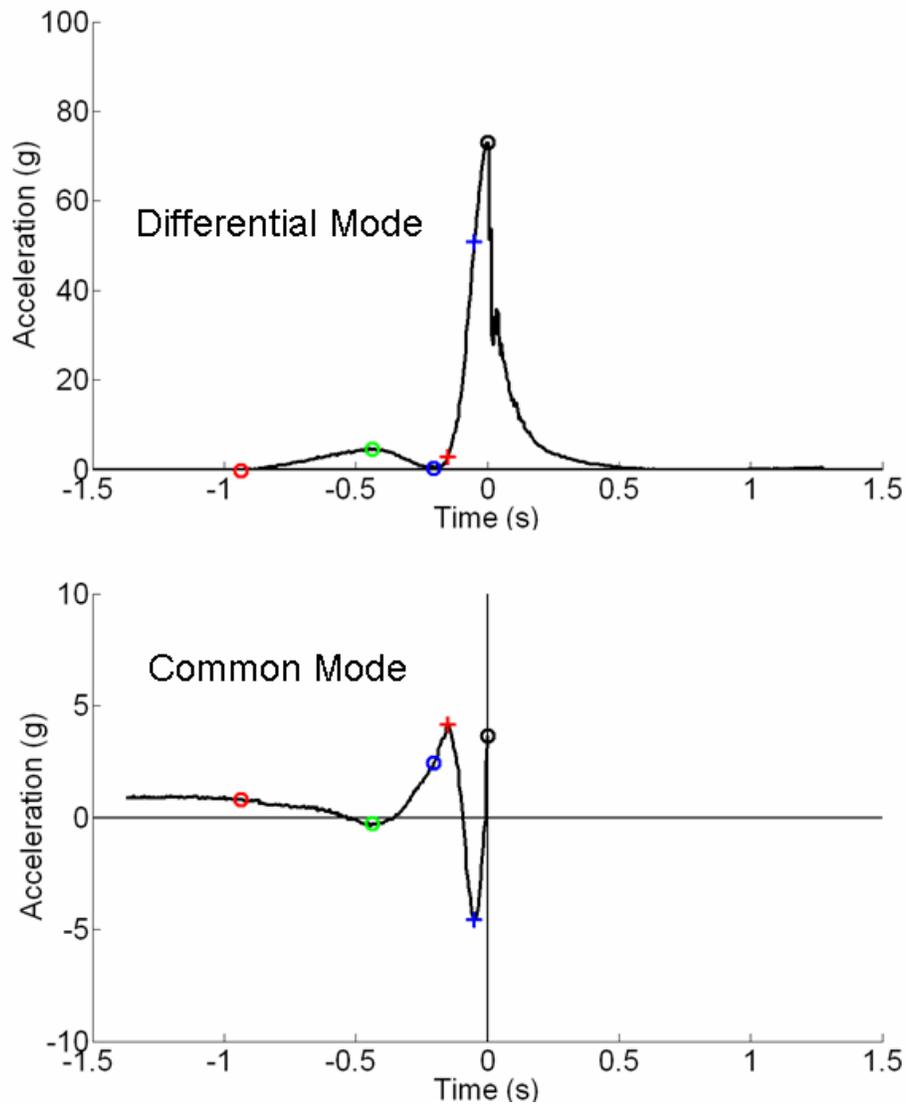

Fig. 5: Differential and common mode data averaged over four swings of a collegiate golfer. The duration of the backswing is 692 ± 48 ms and the duration of the downswing is 201 ± 33 ms, which is extremely fast but not particularly beneficial. In particular, fluctuations in tempo are of order 10%. The club speed at impact is less than that of the professional golfer shown in the following figure. While the common mode data shows a very nicely developed max-min structure, it is smaller in scale than what is shown in Figs. 6 and 7, consistent with the slower club speed.

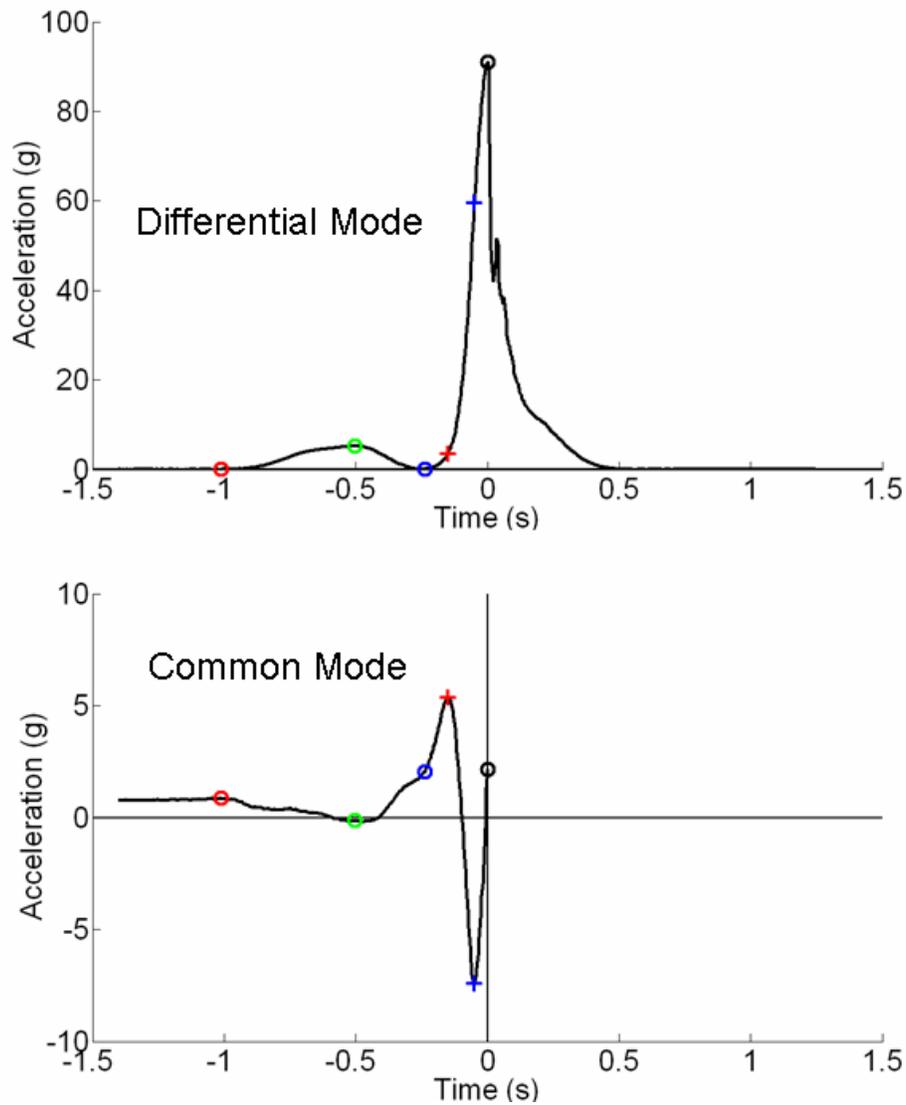

Fig. 6: Differential and common mode data averaged over eight swings of a PGA tour professional. The duration of the backswing is 724 ± 23 ms and the duration of the downswing is 248 ± 4 ms. Note that in comparison with the collegiate golfer the max-min structure in the common mode is larger and wider. This implies that larger torques are sustained for longer periods of time over a larger swing arc.

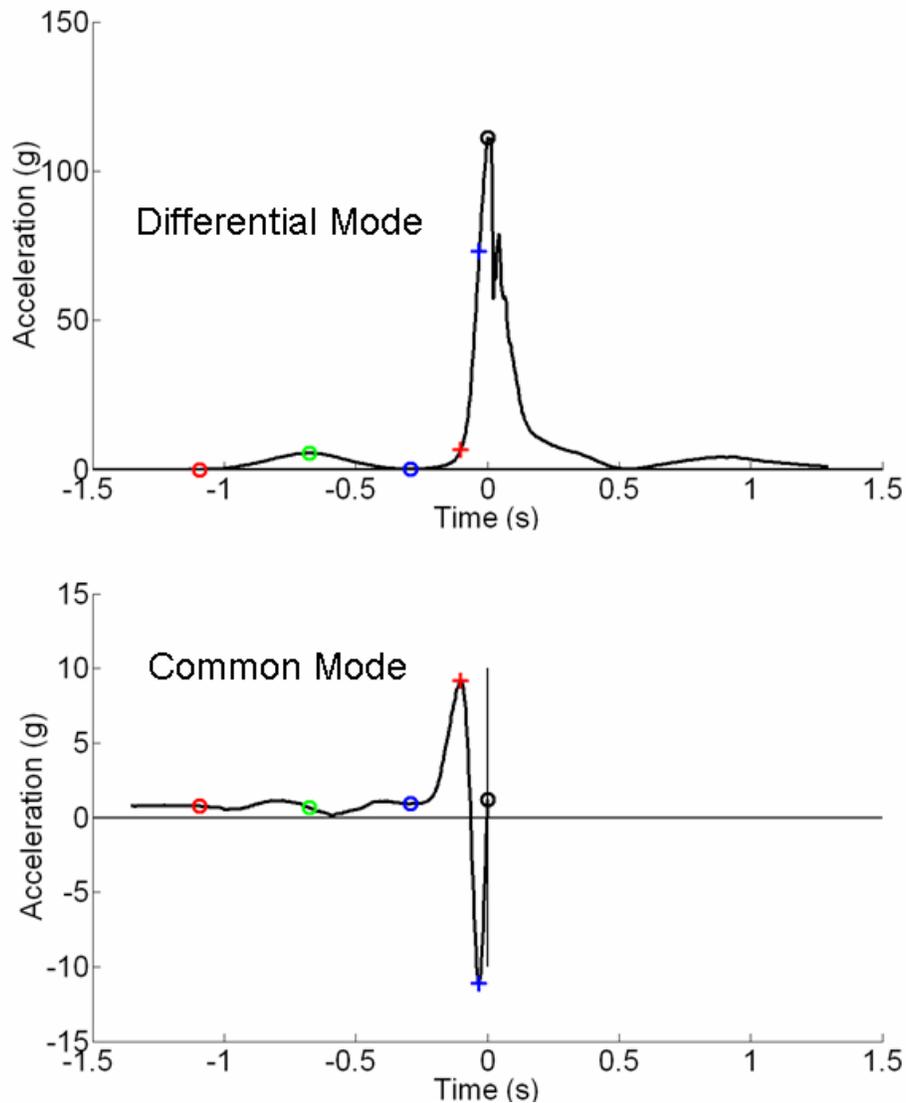

Fig. 7: Differential and common mode data averaged over eight swings of a professional long drive competitor. The duration of the backswing is 750 ± 30 ms and the duration of the downswing is 310 ± 11 ms, though it is clear that the time over which the large torques are applied in the downswing is considerably shorter. Note that **the vertical axes have been increased by 50%** relative to the other golfers in order that that the entire data set can be viewed. This golfer hits the ball far because he is very strong, generating an incredible amount of torque at the beginning of the downswing, with a peak in the common mode that is nearly twice the value for the tour professional.

Comparative Study of 25 Golfers

The implication of the above series of data is that the greater the max-min structure of the common mode signal during the downswing, the greater the resulting club speed. We have tested this hypothesis by comparing the swings of 25 golfers of varying ability, from high-handicapper to tour professional.

The test requires that we compare maximum club speed with the size of the common mode max-min structure. The peak value of the differential mode signal near to impact is used as a proxy for club speed.

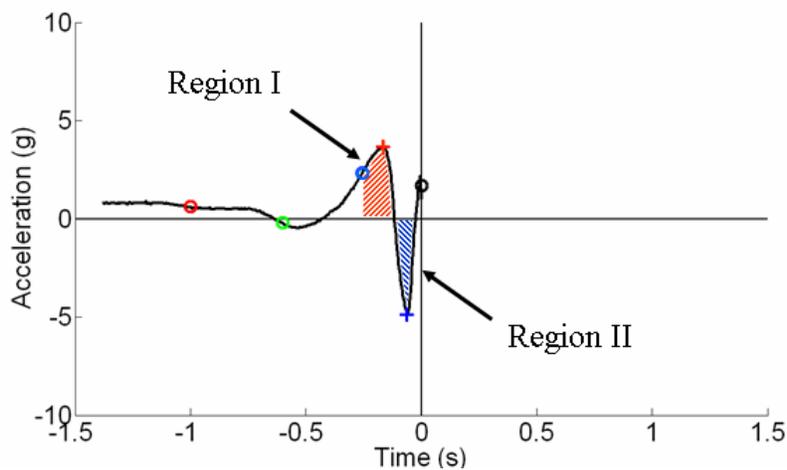

Fig. 8: The common mode signal for the golf swing of a professional golfer. As a measure of the size of the max-min common mode structure during the downswing, the area of two regions is calculated. The first region (red shaded region) starts at the beginning of the downswing (i.e blue circle), goes through the maximum, and terminates at the zero crossing. This area is interpreted as approximating the maximum hand speed during the downswing. The second region (blue shaded region) spans the negative going region of the common mode signal. This area is interpreted as approximating the amount by which the hands slow down as impact is approached.

Our proxy for the size of the common mode max-min structure involves integrating over two regions of the signal, as is shown in Fig. 8. Region I is shown as the red shaded region. It begins at the start of the downswing, a point at which the entire system is moving very slowly. Region I continues through the common mode peak and terminates at the zero crossing. This area is equal to a speed, $v_I$, which is approximately the maximum speed of the hands during the downswing. Region II is indicated as the blue shaded region and spans the entire negative region of the common mode signal. This area yields a negative number, $v_{II}$, which is interpreted as approximating the amount by which the hands slow down as the club is released. Finally, as a measure of the size of the max-min structure $v_I$ is added to the negative of $v_{II}$, yielding $v_F = v_I - v_{II}$.

In Fig. 9, the maximum value of the differential mode signal is plotted relative the integrated common mode signal, $v_F$, for 200 golf swings sampled from 25 golfers. The golfers range widely in capability from high-handicappers to PGA Tour professionals and professional long ball competitors. For each golfer, 5-10 swings are recorded while they are hitting either a 5 or 6 iron. The data indicate a clear trend: *the larger is the common mode max-min structure, the greater the club speed*.

It is important to emphasize that while these measurements are precise, their interpretation is approximate. In particular the interpretation of the areas $v_I$ and $v_{II}$ as a change in hand speed are only accurate in certain limits. For instance, for these speeds to be accurate it must be that the club is moving in the same plane as the hands, the width of the swing does not change during the downswing, the wrists are well cocked throughout the downswing, etc. Indeed, it is likely that none of these constraints are maintained and that the actual conditions vary for every golfer and for every golf swing, which perhaps accounts for the width of the distribution in Fig. 9. As an example, if a golfer lays the club off at the top of the swing, making it flat relative to the path of the hands, the sensors in the shaft are not perfectly aligned with the direction of motion of the hands. In this condition, the common mode maximum will be suppressed.

So, while there is no delusion here regarding perfect interpretation of measurement, the trend in Fig. 9 is clear. The maximum of the differential mode signal, which is a very reasonable proxy for club speed, scales with the size of the common mode max-min structure, which is a very reasonable proxy for the torques that generate club speed.

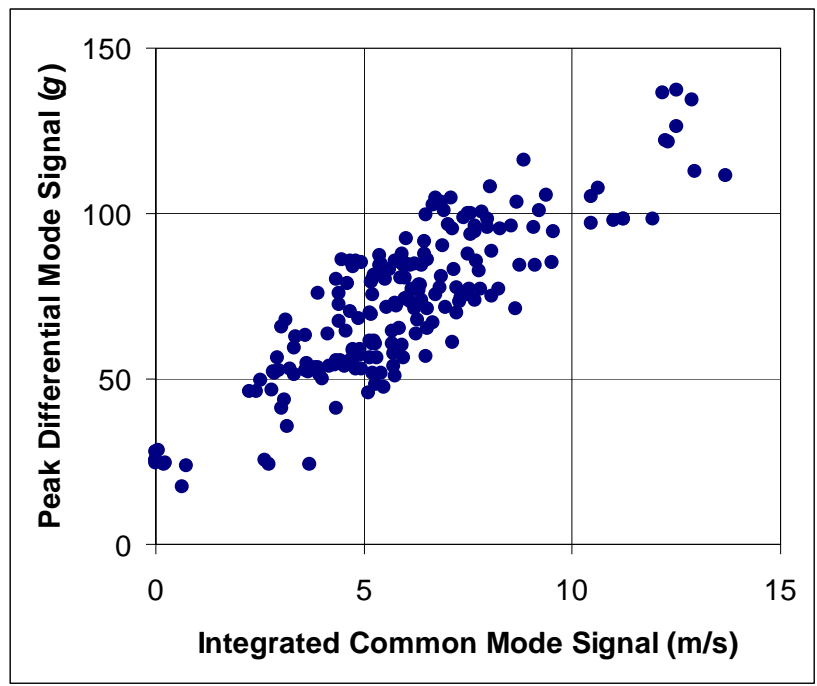

Fig. 9: The peak value of the differential mode signal as a function of the integrated common mode signal, $v_F$, for 200 golf swings sampled from 25 golfers. As defined in the text, $v_F$ is a proxy for the size of the max-min structure in the common mode signal. The peak of the differential mode signal is a proxy for the club speed at impact. The golfers range widely in capability from high-handicappers to PGA Tour professionals and professional long ball competitors. That both signals increase together validates the interpretation of the common mode signal as a measure of the torques which generate speed in the golf swing.

In summary, the common mode signal provides deep insight into how club speed is generated in the golf swing. It is a two step process. Starting at the beginning of the downswing, the first phase involves a rapid acceleration of the hands and club. The height and width of the common mode maximum is a measure of this initial acceleration. This is then followed by a second phase, the release, in which the club accelerates while the hands decelerate. The depth and width of this common mode minimum is a measure of the intensity of the release. Generating club speed requires both phases of acceleration.

Conclusion

The motion of the golf club has been measured using two accelerometers mounted at different points along the shaft of the golf club, both sensitive to acceleration along the axis of the shaft. The resulting signals are resolved into differential and common mode components. The differential mode, a measure of the centripetal acceleration of the golf club, is a reasonable proxy for club speed and can be used to understand details of tempo, rhythm, and timing. The common mode, related to the acceleration of the hands, enables deep insight into the torques that generate speed in the golf swing.

A comparative study of twenty-five golfers reveals that club speed is generated as a two step process. Starting at the beginning of the downswing, the first phase involves a rapid acceleration of the hands and club. This is then followed by a second phase, the release, in which the club accelerates while the hands decelerate.

This paper demonstrates that this measurement scheme provides deep insight into tempo, rhythm, timing, and the torques which generate power in the golf swing.

Footnotes and References:

1) R.D. Grober, *An Accelerometer Based Instrumentation of the Golf Club: Measurement and Signal Analysis*, arXiv:1001.0956v1 [physics.ins-det] .

2  R.D. Grober, *An Accelerometer Based Instrumentation of the Golf Club: Comparative Analysis of Golf Swings*, arXiv:1001.0761v1 [physics.ins-det] .

3)  See United States Patent #7160200 and Sonic Golf, Inc., www.sonicgolf.com , for more information.

4)  J. Novosel and J. Garrity, *Tour Tempo*, (Doubleday, New York, 2004).

5)  R.D. Grober, J. Cholewicki, *Towards a Biomechanical Understanding of Tempo in the Golf Swing*, arXiv:physics/0611291v1.